\documentclass[%
 reprint,
superscriptaddress,
showpacs,
showkeys,
 amsmath,amssymb,
 aip
]{revtex4-2}

\usepackage{lipsum} 
\usepackage{graphicx}
\usepackage{dcolumn}
\usepackage{bm}
\usepackage[breaklinks]{hyperref}
\hypersetup{
  colorlinks,
  citecolor=blue,
  linkcolor=black,
  urlcolor=blue}
\graphicspath{{./PS/}}
\usepackage{tikz}
\usepackage{overpic}
\usepackage{color}
\usepackage{comment}
\usepackage{upgreek}
\usepackage[T1]{fontenc}
\usepackage[utf8]{inputenc}
\usepackage{mathtools}
\usepackage{mathrsfs}
\usepackage[normalem]{ulem}
\usepackage{siunitx}
\DeclareMathOperator{\sgn}{sgn}

\newcommand{\grad}{\bm{\nabla}}

\newcommand{\del}{\nabla^2}
\newcommand{\vct}[1]{{\bf #1}}

\newcommand{\tu}[1]{\mathrm{#1}}

\newcommand{\epsv}{\epsilon_0}
\newcommand{\epsi}{\epsilon_\infty}
\newcommand{\epss}{\epsilon_\mathrm{s}}
\newcommand{\sigc}{\sigma^\mathrm{f}}
\newcommand{\sigp}{\sigma^\mathrm{p}}
\newcommand{\kb}{k_\mathrm{B}}

\newcommand{\Pc}{\mathcal{P}}
\newcommand{\Sc}{\mathcal{S}}

\newcommand{\Fc}{\mathscr{F}}
\newcommand{\Kc}{\mathcal{K}}
\newcommand{\Ac}{\mathcal{A}}

\newcommand{\fp}{$f^*_\tu{DHL}(\kappa_+ \eta;
  \Sc_{1+}^\tu{eff},\Sc_{2+}^\tu{eff})$}
\newcommand{\fm}{$f^*_\tu{DHL}(\kappa_- \eta;
  \Sc_{1-}^\tu{eff},\Sc_{2-}^\tu{eff})$}
\newcommand{\fj}{$f^*_\tu{DHL}(\kappa_j \eta;
  \Sc_{1j}^\tu{eff},\Sc_{2j}^\tu{eff})$}

\begin{document}

\title{Interparticle Interactions in Nonlocal Media: Attraction and Repulsion from Charge-Polarization Coupling}


\author{Ali Behjatian}

\affiliation{Physical and Theoretical Chemistry Laboratory, Department
  of Chemistry, University of Oxford, South Parks Road, Oxford OX1
  3QZ, UK.}

\author{Madhavi Krishnan}%
\email{madhavi.krishnan@chem.ox.ac.uk}

\affiliation{Physical and Theoretical Chemistry Laboratory, Department
  of Chemistry, University of Oxford, South Parks Road, Oxford OX1
  3QZ, UK.}

\affiliation{The Kavli Institute for Nanoscience Discovery, Sherrington Road, Oxford OX1 3QU, UK. }%


\begin{abstract}
Recent measurements of microsphere interactions in diverse media
  suggest that the standard dielectric-continuum models of
  solution-phase interactions are fundamentally
  incomplete. Experiments indicate that the interactions of charged particles in liquids can be dominated by solvent
  structuring at interfaces, thereby motivating the concept of
  electrosolvation.  While interfacial spectroscopy and molecular
  simulations have established that solvent molecules can exhibit net
  orientation at interfaces, conventional theoretical frameworks treat
  the fluid as a structureless medium described by a constant
  dielectric permittivity. This view does not envisage a contribution
  of interfacial polarization to interactions at longer range. Here, we employ nonlocal dielectric theory accounting for
  spatial correlations in polarization to describe interactions in solution. This model permits both charge and polarization to govern
  interactions, leading to dramatic departures from classical
  expectations. Specifically, the balance between charge and
  polarization generates a framework of symmetric (repulsive) and
  antisymmetric (attractive) interactions, wherein: (i) like-charged surfaces
  can attract at long range, (ii) oppositely charged objects can
 repel, and (iii) neutral matter can acquire
  effective electrical mobility and display long-range
  forces—potentially explaining long-range hydrophobic attraction. Further, like-charged biomolecules can attract in aqueous electrolytes
  even for modest polarization correlation lengths ($\xi=2$ \AA). Our results also suggest that electrosolvation effects
  may underpin flocculation in suspended matter,
  which has traditionally been attributed to attractive dispersion
  forces. These findings indicate how
  solvent structuring and correlations may play a dominant, complex
  role in fluid-phase physics, necessitating a shift beyond
  traditional continuum models to accurately describe and understand
  soft matter and biological interactions.
\end{abstract}
                       
\maketitle

\section{Introduction}
Interactions between charged molecules, particles and surfaces in the
fluid phase play a defining role in the emergence of order, structure
and organization in natural systems. Despite significant advances in
the measurement and understanding of forces between surfaces, a
consistent physical picture of the interactions between particle and
molecules in solution remains lacking \cite{israelachvili1978,
  horn1981,israelachvili1982,israelachvili1983}.

Experimental reports over the last several decades have identified a
number of apparently anomalous phenomena that are not readily
explainable within standing theories of interparticle interactions
\cite{langmuir1938,klug1959,kepler1994,grier1998}. A non-exhaustive
list of particularly intriguing experimental effects includes
attraction and cluster formation in negatively charged particles in
water, charge-asymmetric attractive forces that are solvent dependent,
as well as long-range hydrophobic attraction between net neutral
surfaces \cite{eriksson1989,gomez2009,wang2024, wang2025,wang2026}.
Other families of potentially related observations include electrical
mobility of droplets and formally charge-neutral objects in suspension
\cite{marinova1996,egawa1999,schoeler2013,chibowski2016}, as well as
the ability of highly solvated zwitterionic polymers to act as
anti-fouling, highly repellent surface coatings \cite{mi2014}. Both
individually and collectively, these data indicate that the role of
the solvent in governing the interactions of objects—whether with each
other or with external fields—remains rather poorly understood.

The canonical theoretical view of electrostatic interactions treats
the fluid or solvent as a continuum that merely provides a
structureless shielding background \cite{kornyshev1978}. The medium is
described by a phenomenological electric susceptibility constant
${\chi>0}$ and a linear relationship between the polarization,
$\vct{P}$, and the electric field, $\vct{E}$, i.e.,
$\vct{P} = \epsilon_0\chi \vct{E}$, which provides the mathematical
definition of a \textit{local} medium.  Far from a structureless
continuum, however, fluid media are in fact grainy.  The structure of
a fluid is sustained by short-range intermolecular interactions such
as dipole–dipole interactions and hydrogen bonding, but fluids may
also display long-range orientational correlations according to
several experimental reports
\cite{shelton2012,shelton2014,chen2016,duboisset2018,dedic2019,dedic2021}.
Furthermore, molecular simulations \cite{reif2016, loche2018,
  kubincova2020, walker2022} as well as spectroscopy experiments
\cite{du1994,ye2001,shen2006, myalitsin2016} have shown that solvent
molecules display a net orientation or polarization even at neutral
interfaces (Fig. \ref{fig:schm}).  Indeed, the incorporation of a
description of water ordering at interfaces led to successful modeling
of the short-range hydration force which was measured in the 1980s
\cite{marcelja1976,israelachvili1983}.

We recently described a model of electrostatics in a nonlocal medium
capable of sustaining molecular orientational correlations that decay
over a distance $\xi$—the polarization correlation length
\cite{gruen1983a, gruen1983b, behjatian2025}.  We found that depending
on the value of $\xi$, polarization at an interface could make
substantial contributions to the electrical potential at significant
distances. Thus, in addition to the fixed charge density, $\sigc$, at
an interface, the polarization charge density $\sigp$, can determine
whether the particle or surface appears either positively or
negatively charged to an observer at some distance.
Counterintuitively, even charge-neutral objects or surfaces
($\sigc=0$) may appear electrically charged on account of
$\sigp \neq0$.

Here we show that polarization at an interface can have a profound
impact on interactions in solution. We mainly focus on interactions in
a highly polarizable medium like water, containing monovalent ions at
low concentrations (<1 mM). These conditions lie decidedly outside the
strong-coupling regime, and the predictions of
Derjaguin-Landau-Verwey-Overbeek (DLVO) theory are unequivocally
expected to hold \cite{derjaguin1941,verwey1948}. We find that
depending on the balance of fixed charge and polarization charge on
the object, a range of interaction outcomes are possible. We may
expect either attraction or repulsion regardless of whether the
objects are like-charged, oppositely charged or neutral, as well as
non-monotonic interaction potentials displaying minima and maxima,
sometimes entailing multiple reversals in the sign of the
interparticle force.

These results maybe contrasted with the relatively simple view offered
by the DLVO picture which posits that the overall interaction is a
superposition of two monotonic forces: the rather short-ranged
dispersion or van der Waals (vdW) force that is generally always
attractive, and a longer ranged force due to charge, which is either
attractive or repulsive as dictated by traditional Coulombic
expectations. Furthermore, in the DLVO view, the rate of decay of both
the electrostatic potential and force is given by the Debye screening
length $\kappa^{-1} = 1/\sqrt{8 \pi n_0 \ell_\tu{B}}$ where $n_0$
represents the number density of ionic species in the bulk, and
$\ell_\tu{B} = e^2/4 \pi \epsv \epss k_\tu{B}T$ is the Bjerrum length.
Here, the parameters $e$, $\epsv$, $\epss$, $k_\tu{B}$, and $T$ are
the elementary charge, permittivity of free space, static dielectric
constant of the medium, Boltzmann's constant and absolute temperature,
respectively. In contrast, a nonlocal medium is characterized by an
additional solvent-dependent length scale $\xi$ which gives rise to
more than one effective screening length in the decay of the
electrical potential \cite{paillusson2010,behjatian2025}. We further
show that the interaction energy between two objects is characterized
by a hierarchy of decay lengths, the longest of which are amenable to
measurement using microscopy- and scattering-based measurements of
suspension structure.  Importantly, the limit $\xi \to 0$ taken in the
nonlocal model reduces the problem to the traditional
Poisson-Boltzmann (PB) description of a local medium, which then
quantitatively captures the electrostatic component of the DLVO theory
for fluid phase interactions.  Note that the true net interaction in a
nonlocal medium formally includes that due to dispersion forces, but
this contribution is neglected in this study for simplicity.
  
\section{Model}
We consider a system of volume $\Omega$ filled with a nonlocal
electrolyte. The boundary $\partial \Omega$ of the domain, e.g., the surfaces of immersed objects, may carry both
fixed and polarization charges whose surface densities are given by
$\sigc$ and $\sigp$, respectively (Fig. \ref{fig:schm}). The free
energy functional of the system, which depends on the polarization
$\vct{P}$, electrical potential $\phi$, and ionic densities $n_i$, can
be expressed as
\begin{widetext}
\begin{equation}
   \begin{aligned}
    \Fc[\vct{P},\phi, n] =  \int_\Omega \Bigg[ &                                
       -\frac{\epsv \epsi}{2}
        \lvert \grad \phi \rvert^2                                
       + 
        \vct{P}\cdot \grad \phi                                
       + 
       \sum_{i \in \Ac} e q_i n_i \phi
                                + 
       \kb T 
       \sum_{i \in \Ac} \left( n_i \log
       \left(\frac{n_i}{n_i^0}\right) - n_i+ n_i^0
                     \right) \\
       & +
       \frac{\Kc}{2 \epsv \epsi} P^2
       + 
       \frac{\Kc_\ell}{2\epsv \epsi}
       (\grad \cdot \vct{P})^2
       \Bigg]\tu{d}\vct{x}
       + 
         \int_{\partial \Omega} \sigc \phi \tu{d}S
       +
         \int_{\partial \Omega}
         \omega (\vct{P} \cdot \vct{n} -\sigp) \tu{d}S,
  \end{aligned}
  \label{eq:FeCC}
\end{equation}
\end{widetext}
where $\Kc = 1/(\theta -1)$ and $\Kc_\ell = \lambda^2 \Kc$ are
constants. Importantly, in this description $\lambda$ is a
phenomenological length scale which characterizes the nonlocal
properties of the medium, and its physical significance is discussed
later. Here, $\Ac$ represents an index set for the ionic species in
solution, and the parameters $n^0_i$, and $q_i$ describe the bulk
number density and the valency of the $i$th ionic species,
respectively.  For $1:1$ electrolytes we have $\Ac = \{+, -\}$ which
implies ${q_+ = -q_- = 1}$.  We further have $n^0_+ = n^0_- = n_0$
which reflects electroneutrality in the bulk. Further, the parameter
$\theta = \epss/\epsi$ denotes the ratio of the static dielectric
constant $\epss \approx 80$ to that at high frequencies
$\epsi \approx 5$ (Refs. \onlinecite{saxton1952,kornyshevChemSol1985,belaya1987}).
When $\epsi>1$, the polarization field $\vct{P}$ in
Eq. \eqref{eq:FeCC} represents the configurational (orientational)
part of the total polarization field $\vct{P}_\tu{t}$ in the
medium. All remaining contributions to the polarization, e.g., the
electronic part $\vct{P}_\infty$ are described by a local response,
i.e. $\vct{P}_\infty = \epsv \chi_\infty \vct{E}$ where
$\chi_\infty = \epsi-1$.

In Eq. \eqref{eq:FeCC}, the first three terms in the volume integral
denote electrostatic energies due to free and bound charges.  The
fourth term accounts for the mixing entropy of ionic species. The last
two terms in the volume integral capture the nonlocality associated
with polarization correlation in the medium, and are motivated by the
concept of Landau-Ginzburg theory of phase transition \cite{maggs2006,
  paillusson2010}. The first surface integral represents the
electrostatic energy due to fixed charge $\sigc$ on the boundary
$\partial \Omega$, and the second surface integral constrains
$\vct{P}$ on the boundary. Since $\vct{P}\cdot \vct{n}= \sigp$ on
$\partial \Omega$, a fixed polarization surface charge implies a
Dirichlet boundary condition for the normal component of $\vct{P}$ at
the surface. We enforce this condition using the method of Lagrange
multipliers by adding the last integral, and introducing the function
$\omega$ which serves as a Lagrange multiplier.

According to the variational principle, at equilibrium, the functional
$\Fc$ must be stationary with respect to variations in $\vct{P}$,
$\phi$ and $n_i$. It can be shown that the requirement $\delta \Fc =0$
is met if $\vct{P}$, $\phi$ and $n_i$ satisfy partial differential
equations
\begin{equation}
  -\epsv \epsi \del \phi =
  e(n_+-n_-) - \grad \cdot \vct{P},
  \label{eq:poisson}
\end{equation}
\begin{equation}
  \label{eq:P}
  -\epsv \epsi \grad \phi = \Kc \vct{P} -
  \Kc_\ell \grad(\grad \cdot \vct{P}),
\end{equation}
\begin{equation}
  n_\pm = n_0
  \exp\left(\mp \frac{e \phi}{\kb T} \right),
    \label{eq:npm}
\end{equation}
with boundary conditions: (i)
$\epsv \epsi \grad \phi \cdot \vct{n} = \sigc +\sigp$ and (ii)
$ \vct{P} \cdot \vct{n} = \sigp$ (see supplementary information (SI)
for detail).

The boundary condition (ii) above finds no parallel within local
electrostatics theory where polarization $\vct{P}$ depends linearly on
$\vct{E}$ everywhere.  Here, $\sigp$ is not a free parameter and is
determined self-consistently from $\vct{E}$. In particular, at neutral
surfaces, and in the absence of external fields, the local theory
requires an unpolarized state of matter ($\vct{P} = \vct{0}$). This
implies a polarization surface charge $\sigp=0$ on the boundary if
$\vct{P}$ is continuous. In contrast, molecular dynamics (MD)
simulations of the solid-liquid interface have shown that in the
vicinity of surfaces carrying zero fixed charge ($\sigc = 0$) there
exists a nonzero \textit{excess} polarization field
$\vct{P}_\tu{ex}$. This excess polarization can be estimated from the
net interfacial solvent dipole moment density 
\cite{kubincova2020,walker2022}, and has a magnitude
between $0.01$ and $0.1 e$ nm$^{-2}$, as shown in Fig. \ref{fig:schm}c. 
In other words, a thickness ranging
from $10$ \AA~ down to $2$ \AA~ for the interfacial layer of molecules
implies $\vct{P}_\tu{ex} \cdot \vct{n}=\sigp$ values between
approximately $-0.01$ and $-0.1 e$ nm$^{-2}$.  For solvents at
interfaces the excess polarization has been attributed to favourable
H-bonding or other intermolecular interactions of the interfacial
molecules with the solvent half-space, which structures the solvent
and breaks the symmetry in molecular orientation at an
interface. These indications from MD provide a physical justification
and approximate magnitude for $\sigp$ in our present model. By
requiring an extra boundary condition to be satisfied, the nonlocal
electrostatics theory not only readily accommodates interfacial
polarization, but also permits an exploration of the potential
consequences of interfacial polarization propagating into the bulk
medium.

In general, the excess polarization can be viewed as the difference
between the normal components of the total polarization predicted by
nonlocal and local theories, i.e,
$\vct{P}_\tu{ex}\cdot \vct{n} = (\vct{P}_\tu{t}-
\vct{P}_\tu{t}^\tu{loc})\cdot \vct{n}$.  We may establish a connection
between the normal component of excess polarization
$\vct{P}_\tu{ex} \cdot \vct{n}$ and $\sigp$ by simply considering the
boundary conditions of the two theories. In the nonlocal formulation,
noting that $\vct{P}_\infty = \epsv \chi_\infty \vct{E}$ and
$\vct{E}= -\grad \phi$, the boundary condition (i) may be expressed as
$\vct{P}_\infty \cdot \vct{n} = - \chi_\infty (\sigc+\sigp)/\epsi$
which together with (ii) results in
\begin{equation}
    \vct{P}_\tu{t} \cdot \vct{n} = \frac{\sigp}{\epsi}-
    \left(\frac{\epsi-1}{\epsi}\right) \sigc.
\end{equation}
For a local medium with a dielectric constant $\epss$, we also find that 
$\vct{P}_\tu{t}^\tu{loc} \cdot \vct{n} = -\sigc (1 - 1/\epss)$. Therefore 
the normal component of the excess polarization can be expressed as 
\begin{equation}
\label{eq:Pex}
\vct{P}_\tu{ex}\cdot \vct{n} =
\frac{\sigp}{\epsi}+ 
\left(\frac{\theta -1}{\theta}\right)
\frac{\sigc}{\epsi}.
\end{equation}
This implies that for surfaces carrying no fixed charge ($\sigc=0$),
the polarization charge $\sigp$ at the surface is directly related to
the excess polarization, i.e.,
$\vct{P}_\tu{ex} \cdot \vct{n} = \sigp /\epsi$, which we may estimate
by the value of $\vct{P}_\tu{ex}$ inferred from MD simulations. Note
that in molecular simulations using rigid charged-site models of
water, electronic polarizability is generally not included, which
implies $\epsilon_\infty=1$ in Eq. \eqref{eq:Pex}.

\subsection{Two-Field Formulation of Nonlocal Electrostatics}
\label{sec:2ff}
As a direct consequence of Eq. (\ref{eq:P}) which requires $\vct{P}$
to be an irrotational vector, Eqs. (\ref{eq:poisson})-(\ref{eq:npm})
can be written in a computationally more convenient form as follows
(see SI for detail):
\begin{equation}
  -\nabla^2 \psi = \rho,
  \label{eq:poissonPsi}
\end{equation}
\begin{equation}
  -\epsv \epsi \del \phi = \rho -
  \frac{\epsv \epss}{\lambda^2} \phi +
  \frac{\psi}{\lambda^2}.
  \label{eq:phi}
\end{equation}
The boundary conditions are given by: (i)
$\epsv \epsi \grad \phi \cdot \vct{n} = \sigc +\sigp$ and (ii)
$\grad \psi \cdot \vct{n} = \sigc$. Here,
$\rho = -2 n_0 e \sinh (e \phi/\kb T)$ is the charge density due to
the ions in the medium, and $\psi$ is the displacement potential whose
gradient determines the displacement field in our system, i.e.,
$\vct{D} = -\grad \psi$. It is important to note that
Eqs. (\ref{eq:poissonPsi}) and (\ref{eq:phi}) can be derived from an
integro-differential description of nonlocal electrostatics when the
dielectric function is described by a Lorentzian form
\begin{equation}
  \label{eq:kernel}
  \tilde{\epsilon}(k) = \epsi+\frac{\epsi(\theta-1)}{\lambda^2}
  \left( \frac{1}{k^2+\lambda^{-2}} \right)
\end{equation}
in Fourier space \cite{hildebrandt2004,behjatian2025}. This implies that the
functional $\Fc$ constructed at the outset describes the free energy
of a nonlocal system whose dielectric response is captured by a
Lorentzian function given by Eq. (\ref{eq:kernel}). The use of the
Lorentzian approximation to model the dielectric response of water
dates back to early nonlocal theories of electrostatics
\cite{kornyshevChemSol1985,belaya1987}.

Comparing Eq. \eqref{eq:kernel} and the response functions in Refs.
\onlinecite{kornyshevChemSol1985,belaya1987} reveals the relationship
$\lambda = \xi \sqrt{\theta}$ between the phenomenological quantity
$\lambda$ and physical system parameters $\xi$ and $\theta$. Thus,
while $\xi$ is a polarization correlation length in the medium,
$\lambda$ denotes a polarization screening or decay length in our
model.  Note that molecular simulations have since shown that the
dielectric response of water is far more complex, and is in fact not
described by a simple Lorentzian function in Fourier space
\cite{bopp1996,bopp1998,hedley2023, becker2025}.  However this simple
description serves to highlight important qualitative
implications for interactions in nonlocal media.

\section{Problem Setup}
\label{sec:psetup}
\begin{figure}[t!]
  \includegraphics[width=0.45\textwidth,
  trim={0.0cm -0.cm 0cm 0cm},clip]{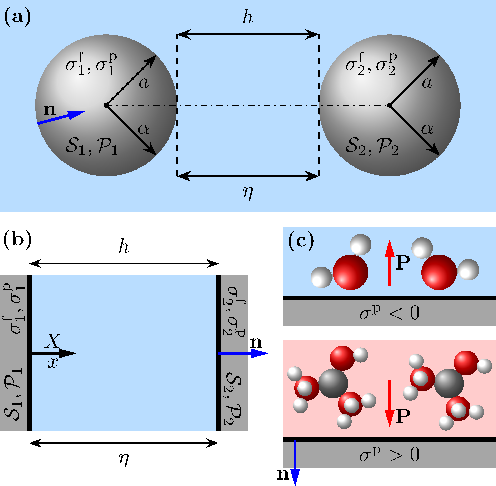}
  \caption{Illustration of systems of interest. (a) Schematic
    representation of two particles of radius $a$ (grey spheres) immersed in a
    nonlocal electrolyte (blue domain). Parameters $\sigc_i$ and $\sigp_i$ represent
    the surface densities of fixed and polarization charges on the
    $i$th particle or plate surface, respectively, with $\Sc_i$ and $\Pc_i$ denoting
    the corresponding dimensionless values.  The unit outward (with respect
    to electrolyte) normal vector to the surface is given by
    $\vct{n}$.  (b) A pair of flat plates carrying fixed and
    polarization charge densities in contact with a nonlocal
    electrolyte. The bottom half of each frame lists system parameters
    and variables in dimensionless form. (c) Elucidation of the
    physical meaning of $\sigp$ in this work. MD simulations of simple
    point charge (SPC) water at a neutral surface ($\sigc=0$) have
    shown that interfacial water molecules possess a small net orientation, which produces a net normal dipole moment
    density or polarization $\vct{P}$ and a corresponding polarization
    charge density $\sigp<0$ (top). At a neutral surface
    immersed in most alcohols, e.g., 2-propanol, interfacial molecules generate a net
    polarization that may be characterized by $\sigp>0$ (bottom).}
 
  \label{fig:schm}
\end{figure}

We use the two-field formulation described in Sec. \ref{sec:2ff} to
model interactions between two flat plates or two spherical particles
of radius $a$ immersed in a nonlocal aqueous electrolyte
(Fig. \ref{fig:schm}). We assume that the charge densities $\sigc_i$
and $\sigp_i$ remain constant as the intersurface separation $h$
between the particles or plates changes.  The thermal voltage
$\phi_0 = \kb T/e$ permits us to write rescaled dimensionless
potentials $\phi_* = \phi/\phi_0$ and
$\psi_* = \psi/\epsv \epss \phi_0$, respectively. Furthermore, we
define the dimensionless position vector $\vct{x} = \kappa \vct{X}$
and gradient operator $\grad_* = \kappa^{-1} \grad$. Equations
(\ref{eq:poissonPsi}) and (\ref{eq:phi}) can then be written in
dimensionless form as
\begin{equation}
  \label{eq:nondPsi}
  -\nabla^2_* \psi_* = \rho_*, 
\end{equation}
\begin{equation}
    \label{eq:nondPhi}
  -\delta^2 \del_* \phi_* = \theta \delta^2 \rho_* +
  \theta (\psi_*-\phi_*),
\end{equation}
for a scaled domain characterized by $\alpha = \kappa a$ and
$\eta = \kappa h$ which represent the dimensionless particle radius
and intersurface separation, respectively (Fig. \ref{fig:schm}). Here,
$\delta = \kappa \lambda$ is the nonlocality parameter and is given by
the ratio of the polarization screening length, $\lambda$, to the
Debye length.  While $\delta \to0$ describes a local medium,
$\delta > 0$ denotes a nonlocal medium. In this work we consider
systems where $\delta$ lies between $0.4$ and $0.8$. Finally,
$\rho_* = - \sinh \phi_*$ represents the dimensionless charge density
due to ions. Accordingly, the boundary conditions for surfaces
$i = 1,2$ transform to
\begin{equation}
  \grad_* \psi_* \cdot \vct{n}_i = \Sc_i
\end{equation}
\begin{equation}
  \grad_* \phi_* \cdot \vct{n}_i = \theta(\Sc_i+\Pc_i).
\end{equation}
Here $\Sc_i = 2 \sgn(\sigc_i)/\kappa \ell^\tu{f}_i$ and
$\Pc_i = 2 \sgn(\sigp_i)/\kappa \ell^\tu{p}_i$ are dimensionless
quantities denoting the fixed and polarization surface charges
respectively. In turn,
$\ell^\tu{f}_i = e/2\pi \lvert \sigc_i \rvert \ell_\tu{B}$ and
$\ell^\tu{p}_i = e/2\pi \lvert \sigp_i \rvert \ell_\tu{B}$ represent
the Gouy-Chapman lengths associated with each particle or surface.  In
all subsequent sections, we work with the dimensionless formulation,
and henceforth, we drop all $*$ symbols for simplicity.
\subsection{Free Energy of The Equilibrium State}
In the $\phi$--$\psi$ formulation, the free energy of the system at equilibrium $\Fc_\tu{eq}$ can be written as
\begin{equation}
  \label{eq:feEq}
  \begin{aligned}
    \frac{\Fc_\tu{eq}}{k_\tu{B}T} = &
    \frac{1}{8 \pi \tau} \Bigg[
    \int_\Omega
    \Big(\phi \sinh \phi-2 \cosh \phi +2 \Big )
    \tu{d} \vct{x} \\
    & 
    + \mu 
    \int_{\partial \Omega} (\phi
      -\psi) \Pc_i \tu{d}S 
    + \int_{\partial \Omega} \Sc_i \phi \tu{d}S \Bigg],
  \end{aligned}
\end{equation}
where $\mu = \theta/(\theta-1)$ and $\tau = \kappa \ell_\tu{B}$ are
two dimensionless constants (see SI for detail). Here, $\Omega$ and
$\partial \Omega$ represent the scaled domain of solution and its
boundary, respectively. In the DH-regime, given by ${\phi \ll 1}$, we
have $\sinh \phi \thicksim \phi$ and
$\cosh \phi \thicksim 1+ \phi^2/2$. The volume integral in
\eqref{eq:feEq} vanishes, and the free energy can be expressed by
\begin{equation}
  \begin{aligned}
    \frac{\Fc_\tu{eq}}{k_\tu{B}T} = &
    \frac{1}{8 \pi \tau} \Bigg[
    \mu
    \int_{\partial \Omega} (\phi
      -\psi) \Pc_i \tu{d}S 
    + \int_{\partial \Omega} \Sc_i \phi \tu{d}S \Bigg],
    \end{aligned}
\end{equation}
where both integrals are defined over the charge carrying
surfaces. For two interacting particles or surfaces, the free energy
$\Fc_\tu{eq}$ is a function of the intersurface separation
$\eta$. Thus we have the dimensionless free energy of interaction
\begin{equation}
\label{eq:F}
    F(\eta) = \frac{\Fc_\tu{eq}(\eta)}{k_\tu{B}T}-
    \frac{\Fc_\tu{eq}(\infty)}{k_\tu{B}T}.
\end{equation}
In this context, $\Fc_\tu{eq}(\infty)$ is the free energy at
infinite separation and the interaction force between the two objects
is given by $-\partial F(\eta)/\partial \eta$ in dimensionless
form.

\subsubsection{Free energy of the parallel-plate interaction}
In one dimension, Eq. \eqref{eq:feEq} simplifies to
\begin{equation}
  \label{eq:fe1dEq}
  \begin{aligned}
    \frac{\Fc_\tu{eq}}{ A_\tu{p} k_\tu{B}T} = &
    \frac{1}{8 \pi A_0} \Bigg[
    \int_0^\eta
      \Big(\phi \sinh \phi-2 \cosh \phi +2 \Big ) \tu{d} x \\
    & 
    + \mu 
    \sum_{i=1}^2 (\phi_i^{\tu{s}}
      -\psi_i^{\tu{s}}) \Pc_i 
    + \sum_{i=1}^2 \Sc_i \phi_i^{\tu{s}} \Bigg],
  \end{aligned}
\end{equation}
where $A_\tu{p}$ is the cross-section area of the flat plates and
$A_0 = \ell_\tu{B} \kappa^{-1}$ is a constant area. Here,
$\phi_i^{\tu{s}}$ and $\psi_i^{\tu{s}}$ represent the surface
potentials of the $i$th surface.  We denote the free energy of
interaction per unit area by
$f(\eta) = [\Fc_\tu{eq}(\eta)-\Fc_\tu{eq}(\infty)]/A_\tu{p}k_\tu{B}T$,
and use $f^*(\eta) = 8 \pi A_0 f(\eta)$ to denote the dimensionless
form of the interaction free energy. As before, in the DH regime, the
integral in Eq. \eqref{eq:fe1dEq} vanishes, and the free energy per
unit area of the interacting surfaces can be obtained from
\begin{equation}
  \label{eq:fe1dEqDH}
  \frac{\Fc_\tu{eq}}{ A_\tu{p} k_\tu{B}T} =
  \frac{1}{8 \pi A_0} \Bigg[
    \mu 
    \sum_{i=1}^2 (\phi_i^{\tu{s}}
      -\psi_i^{\tu{s}}) \Pc_i 
    + \sum_{i=1}^2 \Sc_i \phi_i^{\tu{s}} \Bigg].
\end{equation}

\section{Results and Discussion}
In this section, we investigate interactions between two flat
plates or particles whose fixed and polarization surface charge
densities are given by ($\Sc_1,\Pc_1$) and ($\Sc_2,\Pc_2$),
respectively (Fig. \ref{fig:schm}). We evaluate the free energy of
interaction $F(\eta)$ and quantitatively examine the
distance-dependence of the force -- attractive or repulsive -- for
various combinations of surface properties. To this end, we first
briefly consider the consequences of the nonlocal theory for the
electrical potential profile in the vicinity of an isolated plate.  We
subsequently study the interactions between two parallel plates in the
DH regime (${\phi \ll 1}$), and show how the surface properties of the
two plates defined by ($\Sc_1,\Pc_1$) and ($\Sc_2,\Pc_2$) can give
rise to a variety of interaction energy profiles which are not
obtained within the local PB theory. We then study sphere-sphere
interactions by utilizing two different methods: (i) the Derjaguin
approximation (DA) and (ii) two-dimensional numerical calculations in
an axisymmetric geometry in regimes that may not necessarily satisfy
the DH condition (${\phi \ll 1}$). We will show that in the nonlocal
regime, attractive interactions may be expected for charged and
uncharged colloidal particles in dilute electrolytes
($\kappa^{-1} \approx 100$ nm), as well as for nanospheres such as
proteins at physiological salt concentrations ($\kappa^{-1} \approx 1$
nm).
\subsection{Electrical potential due to a single flat plate}
\label{sec:1dpot}
The potential distribution due to an isolated flat plate carrying
surface charges $\Sc$ and $\Pc$ immersed in a nonlocal electrolyte has
been studied extensively in Ref. \onlinecite{behjatian2025} and the results are briefly recapitulated here. In the DH regime
($\phi \ll 1$), the solution of Eqs.
\eqref{eq:nondPsi}-\eqref{eq:nondPhi} for a one-dimensional
semi-infinite domain has the form
\begin{equation}
\label{eq:potSemiAx}
\phi(x) = A_+ \exp(-\kappa_+ x) + A_- \exp(-\kappa_- x),
\end{equation}
where $A_\pm$ are two coefficients which depend on $\Sc$ and $\Pc$ 
(See Appendix \ref{sec:appPot}). 
Here,
\begin{equation}
  \label{eq:kappa}
  \kappa_\pm =
  \left[\frac{\theta (1+\delta^2)}{2\delta^2}
    \left( 1 \pm
      \sqrt{1-\frac{4\delta^2}{\theta
          (\delta^2+1)^2}} \right)
    \right]^{1/2},
\end{equation}
are screening parameters that denote dimensionless screening lengths.
Since $\kappa_+/\kappa_->\sqrt{\theta}\gg1$, the behavior of $\phi(x)$
at large distances $x$ is governed by $\kappa_-$ 
(See Ref. \onlinecite{behjatian2025}).

Importantly, the nonlocal model gives rise to more than one screening
length in the decay of the electrical potential.  When $\theta \gg 1$,
the two decay lengths maybe be given in dimensional form by the
following approximations 
\begin{equation}
  \kappa_\tu{S}^{-1} = (\kappa \kappa_+)^{-1} \sim
  \frac{1}{\sqrt{\theta(\lambda^{-2}+\kappa^{2})}}
\end{equation}
and 
\begin{equation}
  \kappa_\tu{L}^{-1} = (\kappa\kappa_-)^{-1}
  \sim {\sqrt{\lambda^2+\kappa^{-2}}}.
\end{equation}
These expressions emphasize the emergence of new effective screening
lengths (short - S, and long -- L) from the explicit coupling between
the two length scales of the problem: (1) the traditional Debye length
$\kappa^{-1}$ associated with screening by ions, and (2) the
polarization screening length $\lambda=\xi\sqrt{\theta}$, which is
inherent to the nonlocal properties of the electrolyte.

In particular we showed that the electrical potential at a large
distance from the surface could be well approximated by
$\phi(x)=\phi_s \exp(-\kappa_\tu{eff} x)$, where $\phi_\tu{s}$ is an
effective potential extracted from a fit to the far-field of the
numerically calculated potential. In physical terms, $\phi_\tu{s}$ 
is the value of the surface potential
from the perspective of a distant observer.  Based on this definition,
the coefficient $A_-$ in Eq. \eqref{eq:potSemiAx} precisely determines
$\phi_\tu{s}$, the effective surface potential, in the DH regime.

\begin{figure}[t!]
  \includegraphics[width=0.48\textwidth,
  trim={0.0cm -0.0cm 0cm 0cm},clip]{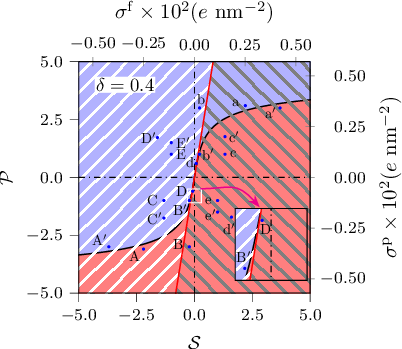}
  \caption{Sign of the effective surface potential $\phi_\tu{s}$ as a
    function of dimensionless charge densities $\Sc$ and $\Pc$ for
    $\delta=0.4$. The inclined red line line represents
    $\phi_\tu{s}^\tu{DH}(\Sc,\Pc)=0$ based on the DH solution. The DH
    solution divides the $\Sc$-$\Pc$ plane into two hatched regions:
    $\phi_\tu{s}^\tu{DH}(\Sc,\Pc)>0$ (gray) and
    $\phi_\tu{s}^\tu{DH}(\Sc,\Pc)<0$ (white). The black curve denotes
    $\phi_\tu{s}(\Sc,\Pc)=0$ based on numerical calculations. Blue and
    red shading indicate regions of the $(\Sc,\Pc)$ parameter space
    where $\phi_\tu{s}<0$ and $\phi_\tu{s}>0$, respectively. Values
    noted for $\sigc$ and $\sigp$ correspond to a dilute electrolyte
    with a Debye screening length of $\kappa^{-1} =100$ nm, as given
    by the relations $\Sc= 4\pi\sigc\ell_\tu{B}/e\kappa$ and
    $\Pc= 4\pi\sigp\ell_\tu{B}/e\kappa$. Lowercase letters represent
    reflections through the origin of points labeled in capital
    letters, e.g., $\tu{a}'$ and $\tu{A}'$. Potential profiles corresponding to reflection-symmetric points differ only in sign, i.e., their $\lvert \phi(x)\lvert$ profiles are identical.}
  \label{fig:phaseMap}
\end{figure}

\begin{figure}[t!]
  \includegraphics[width=0.45\textwidth,
  trim={0.0cm -0.0cm 0cm 0cm},clip]{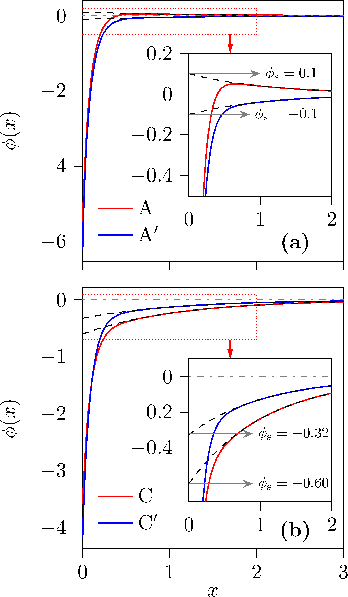}
  \caption{ Spatial profiles of electrical potential $\phi(x)$ for an
    isolated flat plate in contact with a nonlocal medium. Solid lines
    in (a) and (b) display potential profiles for pairs of points in
    Fig. \ref{fig:phaseMap} that can give rise to antisymmetric
    ($\tu{A}$ and $\tu{A'}$) and symmetric ($\tu{C}$ and $\tu{C'}$)
    parallel-plate interactions, respectively.  Dashed black lines
    represent fits of the form
    $\phi(x)=\phi_\tu{s} \exp(-\kappa_\tu{eff} x)$ to the numerically
    calculated potentials, which yield the map presented in
    Fig. \ref{fig:phaseMap}}.
  \label{fig:potDist}
\end{figure}

In stark contrast to a local medium, the sign of the effective surface
potential $\phi_\tu{s}$ in a nonlocal electrolyte does not solely
depend on the sign of the fixed charge $\Sc$ at the interface. In
fact, depending on the value of a nonlocality parameter $\delta$, the
effective surface potential $\phi_\tu{s}$ is given by a two-variable
function of $\Sc$ and $\Pc$. Furthermore, the curve defined by
$\phi_\tu{s}(\Sc,\Pc) = 0$ divides $\Sc$-$\Pc$ plane into two distinct
regions of positive and negative values of effective surface
potentials (Fig. \ref{fig:phaseMap}). Thus the sign of $\phi_\tu{s}$
distinguishes the two regions $\phi_\tu{s}(\Sc,\Pc)<0$ (blue) and
$\phi_\tu{s}(\Sc,\Pc)>0$ (red) in Fig. \ref{fig:phaseMap}.  The
solution in the DH regime also divides the $\Sc$-$\Pc$ plane into two
distinct regions of positive and negative surface potentials (hatched
regions in Fig. \ref{fig:phaseMap}). However, the curve or contour
defined by $\phi_\tu{s}^{\tu{DH}}(\Sc, \Pc)=0$ and
$\phi_\tu{s}(\Sc, \Pc)=0$ are markedly different from one another for
large values of $\Sc$ and $\Pc$, where ${\phi \ll 1}$ is
violated. Figure \ref{fig:potDist} displays the spatial variation of
electrical potential for pairs of points $(\tu{A},\tu{A}')$ and
$(\tu{C},\tu{C}')$ together with exponential fits used to extract
$\phi_\tu{s}$. Note that for two points, e.g., $\tu{A}$ and $\tu{A}'$
with negative values of both fixed surface charge and polarization charge values,
($\Sc_\tu{A}, \Pc_\tu{A}<0$ and $\Sc_\tu{A'}, \Pc_\tu{A'}<0$), the
effective surface potentials can be of opposite sign. 
A similar result has been obtained for surfaces carrying both charge and dipole moment density, immersed in a local medium \cite{belaya1994a}. 

For fixed $\delta$ and $\theta$, a single plate in a nonlocal medium
can be geometrically identified by a single point in two-dimensional
Euclidean space $\mathbb{R}^2$. We use this idea to systematically
describe the interactions of parallel-plate and two-sphere
systems. Once the parameters $\delta$ and $\theta$ are fixed,
interactions are fully specified by choosing a pair of points
$(\Sc_1, \Pc_1)$ and $(\Sc_2, \Pc_2)$ in $\mathbb{R}^2$. Based on this
geometric view, it is evident that for weakly interacting flat plates
(${\eta \gg 1}$) with $(\Sc_1, \Pc_1) = (\Sc_\tu{A}, \Pc_\tu{A})$ and
$(\Sc_2, \Pc_2) = (\Sc_\tu{A'}, \Pc_\tu{A'})$, the potential $\phi(x)$
must go through zero at some point between the plates. In a local
medium, this situation only occurs if the two surfaces are oppositely
charged. Moreover local PB models have shown that the force between
two oppositely charged surfaces with unequal charge densities
($\Sc_1 \neq \Sc_2$) is always attractive over long distances and
turns repulsive over short distances as the repulsive entropic
contribution dominates the interaction
\cite{parsegian1972,mccormack1995}.  This analogy immediately suggests
the possibility of a long-range attractive force between surfaces with
effective surface potentials of opposite signs, i.e.,
$\phi_\tu{s}(\Sc_1, \Pc_1) \phi_\tu{s}(\Sc_2, \Pc_2) <0$. For
convenience, we call such systems \textit{antisymmetric}. We consider
a system \textit{symmetric} if the $\phi_\tu{s}$ values of the
interacting entities are of the same sign, i.e.,
$\phi_\tu{s}(\Sc_1, \Pc_1) \phi_\tu{s}(\Sc_2, \Pc_2) >0$.

\subsection{Interaction Free Energy for Flat Plates}
\label{sec:DH}
The dimensionless free energy of interaction between two flat plates
with charge densities $\Sc_1$ and $\Sc_2$ in a local medium
($\delta \to 0$) in the DH regime can be expressed as
\begin{equation}
  \label{eq:FE1dDHL}
  \begin{aligned}
    f^*_\tu{DHL}(\eta;\Sc_1,\Sc_2) = &
  \Bigg [
        (\Sc_1^2+\Sc_2^2) (\coth \eta -1)+ \\
    &   \frac{2 \Sc_1 \Sc_2}{\sinh \eta}
       \Bigg ],
  \end{aligned}
\end{equation}
where $\Sc_1^2+\Sc_2^2>0$ and $\Sc_1 \Sc_2$ are two constants which
determine the overall behavior of
$f^*_\tu{DHL}(\eta;\Sc_1,\Sc_2)$. The function $f^*_\tu{DHL}(\eta)$ is
a monotonically decreasing function of $\eta$ if $\Sc_1 \Sc_2>0$. When
$\Sc_1 \Sc_2 <0$ however, $f^*_\tu{DHL}(\eta)$ exhibits non-monotonic
behavior with a unique minimum at
$\eta_\tu{min} = \ln(\sqrt{1/a_0^2-1}-1/a_0)$, where
$a_0 = 2 \Sc_1 \Sc_2/(\Sc_1^2+\Sc_2^2)$ and $\lvert a_0 \rvert <1$. In
the nonlocal regime, it can be shown that the free energy of
interaction is described by the combination of two similar functions
\begin{equation}
  \label{eq:Fe1dDH}
  f^*_\tu{DH}(\eta) = \sum_{j \in \mathcal{A}}   
  f^*_\tu{DHL}(\kappa_j \eta; \Sc_{1j}^\tu{eff},
  \Sc_{2j}^\tu{eff})                    
\end{equation}
where each function is governed by one of the screening parameters
$\kappa_\pm$ given by Eq. \eqref{eq:kappa}. In this context, the
parameters $\Sc_{i+}^\tu{eff} = \Omega_+ u_i$ and
$\Sc_{i-}^\tu{eff} = \Omega_- v_i$ describe the effective charges of
the $i$th plate where $\Omega_\pm$ are positive constants (See
Appendix~\ref{sec:appEffCharge}). Here, $u_i = \Pc_i +n \Sc_i$ and
$v_i = \Pc_i -m \Sc_i$ are linear combinations of $\Sc_i$ and $\Pc_i$
with $m=(\kappa_+^2/\theta -1)$ and $n=(m+1)/(\mu m+1)$ being positive
constants.

Examination of the coefficients $A_\pm$ (Eq. \eqref{eq:potSemiAx}, Appendix \ref{sec:appPot})
suggests that a re-parametrization of $\mathbb{R}^2$ based on the
variables $u=\Pc+n\Sc$ and $v=\Pc-m \Sc$ provides a more effective
description of interactions in a flat plate system. The $u$-$v$
coordinate system may be justified by first noting that the curve
$\phi^\tu{DH}_\tu{s}(\Sc, \Pc)=0$, describing the loci of points at
which $A_-=0$, coincides with the line $v=0$ ($u$-axis)
[Fig. \ref{fig:dhCoords}a top].  Similarly, the loci of points at which
$A_+ = 0$ determines a line $\Pc = -n \Sc$ which is identical to $u=0$
($v$-axis) [Fig. \ref{fig:dhCoords}a top]. Accordingly, we infer that the
$u$- and $v$-axes identify special points in the $\Sc$-$\Pc$ plane at
which the potential away from the surface is described by a single
exponential function albeit with a different effective screening
parameter, either $\kappa_+$ or $\kappa_-$.  For polar liquids where
${\theta \gg 1}$, the coefficient $n$ is close to unity, and therefore
the line $\Pc = -n \Sc$ forms approximately equal angles with $\Sc$-
and $\Pc$-axis in the second and fourth quadrants in the $\Sc$-$\Pc$
plane. Defining the coordinate transformations $u = \Pc + n \Sc$ and
$v = \Pc-m \Sc$, we observe that for all $\delta>0$ and $\theta>1$,
there is a one-to-one map between $(u,v)$ and $(\Sc, \Pc)$
coordinates. Parallel-plate systems identified by a pair of points
$(\Sc_1, \Pc_1)$ and $(\Sc_2, \Pc_2)$ in the $\Sc$-$\Pc$ plane are now
equivalently described by $(u_1, v_1)$ and $(u_2, v_2)$ in the $u$-$v$
plane, where $u_i = \Pc_i +n \Sc_i$, $v_i = \Pc_i -m \Sc_i$, and
$i = 1,2$.  As shown in Fig. \ref{fig:dhCoords}a, the $u$- and
$v$-axis are, in general, not orthogonal to each other in the
$\Sc$-$\Pc$ plane; rather, they divide the plane into 4 distinct
regions ($\tu{A}$, $\tu{B}$, $\tu{C}$, and $\tu{D}$) whose images
under the transformation correspond to perfect quadrants in the
$u$-$v$ plane (Fig. \ref{fig:dhCoords}a bottom). The $(u,v)$ coordinate
system enables a systematic exploration of the various interaction
regimes in a nonlocal medium (Fig. \ref{fig:dhCoords}).

Finally we note that Eq. \eqref{eq:Fe1dDH} greatly simplifies the
inference and classification of overall trends. It implies that the
interaction of two plates immersed in a nonlocal electrolyte can be
viewed as the sum of interactions of two parallel-plate
subsystems. Importantly, the plates in each case are immersed in a
hypothetical \textit{local} electrolyte which is characterized by
either one of the screening parameters $\kappa_+$ or $\kappa_-$.
\subsubsection{Interaction regimes for flat plates}
Here, we establish general conditions for the existence of spatially
non-monotonic interactions within the framework of our nonlocal
electrostatics model. We begin with an asymptotic analysis of the
interaction free energy given by Eq. \eqref{eq:Fe1dDH}.  For
sufficiently small separations between the two surfaces
(${\kappa_\pm \eta \ll 1}$), we have
\begin{equation}
\label{eq:FEzeroAsym}
f^*_\tu{DH}(\eta) \sim c \left[
  b \left(\frac{1+a_-}{\kappa_-} \right)+
  \left(\frac{1+a_+}{\kappa_+} \right)
\right] \frac{1}{\eta},
\end{equation}
where
\begin{equation}
  \label{eq:a+}
a_+ = \frac{2 u_1 u_2}{u_1^2+u_2^2},
\end{equation}
\begin{equation}
    \label{eq:a-}
a_- = \frac{2 v_1 v_2}{v_1^2+v_2^2},
\end{equation}
and $b=[\Omega_-^2(v_1^2+v_2^2)]/[\Omega_+^2(u_1^2+u_2^2)]$ and
$c = \Omega_+^2(u_1^2+u_2^2)$ are coefficients in $u$-$v$
coordinates. It immediately follows that $b$ and $c$ are positive and
${|a_\pm|<1}$. Consequently, the asymptotic relation
\eqref{eq:FEzeroAsym} implies that for sufficiently small intersurface
separations $f^*_\tu{DH}(\eta)$ is strictly positive. 

On the other hand, when ${\kappa_\pm \eta \gg 1}$, noting that
$\coth \kappa_\pm \eta-1 \sim 2 \exp(-2 \kappa_\pm \eta)$ and
$1/\sinh \kappa_\pm \eta \sim 2 \exp(-\kappa_\pm \eta)$, we find that
\begin{equation}
  \label{eq:FEinfAsym}
  \begin{aligned}
  f^*_\tu{DH}(\eta) \sim & 2c \Bigg[
    \exp(-2 \kappa_+ \eta) + a_+ \exp(- \kappa_+ \eta)+\\
    & b \exp(-2 \kappa_- \eta)+ b a_- \exp(\kappa_- \eta)
    \Bigg].
  \end{aligned}
\end{equation}
Furthermore, since ${\kappa_+>\kappa_-}$, the sign of
$f^*_\tu{DH}(\eta)$ at sufficiently large intersurface separations
$({\kappa_- \eta \gg 1})$ is completely determined by that of
$a_-$. This means that if ${a_-<0}$ then $f^*_\tu{DH}(\eta)$ must
change sign from positive to negative as $\eta$ increases. Since by
definition $F(\eta)$ settles to zero at infinite separation, we can
expect that ${a_-<0}$ guarantees the existence of at least one local
minimum.  We find that in most cases, if there is a local minimum when
$\kappa_-\eta \gg 1$ it must be determined by the competition between
the last two terms in Eq. \eqref{eq:FEinfAsym}.  Thus at long range,
Eq. \eqref{eq:FEinfAsym} may be approximated and written in
dimensional form as
\begin{equation}
  \label{eq:FEinfAsym2}
  \begin{aligned}
  f_\tu{DH}(x) \sim &A \exp(-\kappa_1x)+ B\exp(\kappa_2x),
  \end{aligned}
\end{equation}
where $\kappa_1=2\kappa_{\tu{L}}$ and $\kappa_2=\kappa_{\tu{L}}$
indicate the two longest screening lengths in the interaction free
energy. We have previously shown that
the above equation indicates that a
minimum may arise from the superposition of two exponentially decaying
terms, provided $A>0$ and $B<0$ (Ref. \onlinecite{behjatian2022}). Further, within the present nonlocal
model, the inverse screening length $\kappa_{\tu{L}}< \kappa$ governs
both long-ranged terms in the free energy. For a correlation length
$\xi=10$ nm, Debye length $\kappa^{-1}=100$ nm, and $\theta=16$, we
have $\kappa_2^{-1}=\kappa_{\tu{L}}^{-1}\approx108$ nm. Note that
here, $\kappa_{\tu{S}}^{-1}\approx 9$ nm is a much shorter decay
length and influences the system behavior on significantly closer
approach.

\begin{figure*}[t!]
    \centering
    \includegraphics[width=\textwidth]{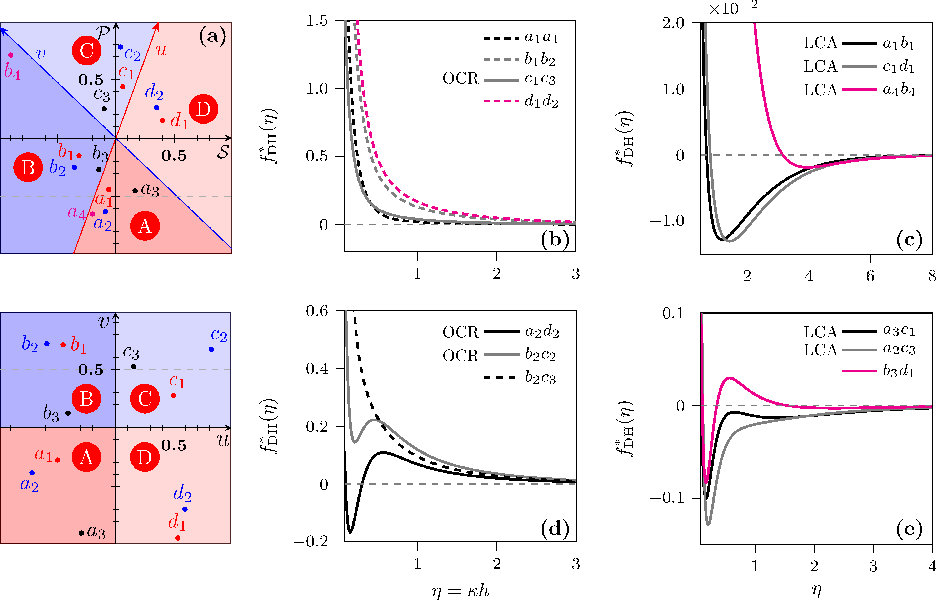}
    \caption{Regimes of interaction for two parallel plates in a
      nonlocal medium in the DH regime ($\delta = 0.6$).  In all free
      energy profile plots, dashed curves denote cases where the
      interaction is qualitatively in line with traditional PB theory,
      whilst solid lines denotes departures from the standard
      expectation due to nonlocal electrostatics.  (a) Top:
      Two-dimensional plots of the sign of the effective surface
      potential $\phi_\tu{s}$ in the $\Sc$-$\Pc$ plane. The $u$-axis
      ($\phi_\tu{s}(\Sc,\Pc) \equiv A_-=0$) divides the $\Sc$-$\Pc$
      plane into regions with positive (red) and negative (blue)
      effective surface potentials, respectively. The $v$-axis denotes
      the loci of points at which $A_+$ vanishes, and together with
      the $u$-axis divides the $\Sc$-$\Pc$ plane into four regions A,
      B, C, and D.  Bottom: Plot of the sign of $\phi_\tu{s}$ in the
      $u$-$v$ coordinate system. Regions A, B, C and D in the
      $\Sc$-$\Pc$ plane correspond to perfect quadrants in the $u$-$v$
      plane.  (b) \textit{Symmetric} interactions between pairs of
      points (plates) located in the same region (either A, B, C or
      D), corresponding to interaction coefficients ${a_\pm>0}$
      (purely repulsive interactions). The curve $c_1c_3$ denotes an interaction between surfaces with properties given by $c_1$ and $c_3$ and highlights a case of opposite-charge repulsion (OCR).  (c) \textit{Antisymmetric} interactions result from
      pairings of surfaces in regions A and B (C and D) corresponding
      to interaction coefficients ${a_-<0}$ and ${a_+>0}$, giving rise
      to a nonmonotonic interaction regime with a unique local
      minimum, characteristic of like-charge attraction (LCA). (d)
      \textit{Symmetric} interactions between pairs of points (plates)
      in regions A and D (B and C), corresponding to coefficients
      ${a_->0}$ and ${a_+<0}$, yield nonmonotonic OCR
      behavior ($a_2d_2$ and $b_2c_2$). (e) Interactions between surfaces in regions A and C (B and D), where $a_\pm <0$,
      yield multiple stationary points at smaller intersurface
      separations. When the Debye length $\kappa^{-1} = 100$ nm,
      $\Sc =\Pc= 0.5$ corresponds to a charge density of
      $5.68\times10^{-4} e$ nm$^{-2}$. Similarly, the dimensionless
      free energy $f_\tu{DH}^*(\eta)=1$ is equivalent to a free energy
      per unit area of $5.68\times10^{-4} k_\tu{B}T$ nm$^{-2}$. The
      regimes of interaction (b)-(e) can be explored interactively
      using a Mathematica application that is available as
      Supplementary Material.  }
    \label{fig:dhCoords}      
\end{figure*}
We discuss four different interaction regimes based on the properties
of the function $f^*_\tu{DHL}(\eta;\Sc_1, \Sc_2)$ discussed
earlier. It is important to recall that the function \fj~in
Eq. \eqref{eq:Fe1dDH} is a monotonically decreasing function of $\eta$
when ${\Sc^\tu{eff}_{1j}\Sc^\tu{eff}_{2j}>0}$ but indicates
nonmonotonic behavior if ${\Sc^\tu{eff}_{1j}\Sc^\tu{eff}_{2j}<0}$. We
also remark that functions \fp~and \fm~govern the short- and
long-range interactions in our systems, respectively. Since the sign
of $\Sc^\tu{eff}_{1j}\Sc^\tu{eff}_{2j}$ is the same as that of $a_j$
given by Eqs. \eqref{eq:a-} and \eqref{eq:a+}, the nature of the
interaction $f^*_\tu{DH}(\eta)$ can be classified based on the signs
of $a_+$ and $a_-$ alone. Thus, based on the long-range/short-range
behavior in Eq. \eqref{eq:Fe1dDH}, there are four possibilities to
consider: (i) repulsion/repulsion (${a_\pm>0}$), (ii)
nonmonotonic/repulsion (${a_-<0}$ and ${a_+>0}$), (iii)
repulsion/nonmonotonic (${a_->0}$ and ${a_+<0}$), (iv)
nonmonotonic/nonmonotonic (${a_\pm<0}$). We explore these interactions
by considering different pairs of points $p_1=(u_1, v_1)$ and
$p_2=(u_2, v_2)$ in the $u$-$v$ plane (Fig. \ref{fig:dhCoords}a bottom).

It follows from the definition that ${a_\pm>0}$ if the points $p_1$
and $p_2$ both lie in the same quadrant, either A, B, C, or
D. Accordingly, the nonlocal interaction of such \textit{symmetric}
systems is equivalent to the superposition of two purely repulsive
\textit{local} interactions. A special case of this result is that if
two surfaces have identical properties, i.e., $\Sc_1 = \Sc_2$ and
$\Pc_1 = \Pc_2$, their interaction is always repulsive.  Thus, for all
symmetric interactions corresponding to ${a_\pm>0}$, the overall
interaction is repulsive at all
separations. (Fig. \ref{fig:dhCoords}b).
 
Interesting trends emerge for interactions between systems given by
points $p_1$ and $p_2$ located in different quadrants of
Fig. \ref{fig:dhCoords}a. Turning our attention to the second scenario
(${a_-<0}$ and ${a_+>0}$), we observe that this condition portrays
\textit{antisymmetric} systems in which the long-range attractive
contribution of \fm~is counterbalanced by the short-range purely
repulsive interaction due to \fp. This situation leads to the
formation of a unique minimum whose location depends on the relative
strength of these interactions. In general, the minimum becomes
shallower and shifts to larger values of $\eta$ as the relative
strength of \fm~decreases. This is because a stronger shorter ranged
repulsive force may overcome the weak attractive contribution of
\fm~at larger intersurface separations (Fig. \ref{fig:dhCoords}c).

Figure \ref{fig:dhCoords}d represents \textit{symmetric} interactions
where the opposite set of conditions holds, i.e., ${a_->0}$ and
${a_+<0}$. Here, the overall behavior of the system is determined by a
long-range repulsive force, which may be overcome by an attractive
contribution from ${a_+<0}$ at shorter range. If the short-range
attraction is weak relative to the repulsive interaction, the overall
interaction may in fact be purely repulsive at all separations
($b_2c_3$ in Fig. \ref{fig:dhCoords}d). But if there exists an
intersurface separation at which the short-range and long-range
contributions are of comparable magnitude, the overall interaction may
exhibit a maximum. We note that a local maximum in the free energy
curve is always accompanied by a minimum which occurs at shorter
separations. This short-range minimum occurs on account of
Eq. \eqref{eq:FEzeroAsym} which holds when $\eta \to 0$. As a result,
both long-range and short-range contributions turn repulsive at
sufficiently small intersurface separations, leading to purely
repulsive interactions as $\eta \to 0$ ($a_2d_2$ and $b_2c_2$ in
Fig. \ref{fig:dhCoords}d).

Finally, ${a_\pm<0}$ represents the situation where both short- and
long-range forces are non-monotonic. The superposition of two
nonmonotonic functions may give rise to interaction free energy
profiles which may exhibit multiple stationary points
(Fig. \ref{fig:dhCoords}e). The behavior of these antisymmetric
systems at long-range, however, closely resembles that of systems
classified under the second case (${a_-<0}$ and ${a_+>0}$).
Systems depicted in Figs. \ref{fig:dhCoords}c and 
\ref{fig:dhCoords}d  are capable of giving rise to shallow local minima at long
range and may therefore be viewed as potential prototypes which may
explain the experimentally observed phenomenon of like-charge
attraction. The regimes of interaction discussed above can also be
explored interactively using a Mathematica application that is
available as Supplementary Material.
  
Overall, these considerations suggest the possibility of multiple
changes in the sign of force as two non-identical objects approach
each other in a nonlocal electrolyte. In particular, we note the
possibility of long-range secondary minima of tuneable depth and range
depending on particle properties. Note that this model takes a purely
electrical view of the problem and steric and dispersion-force contributions at molecular-scale
separation between the surfaces are not considered
\cite{israelachvili1983,schiby1983a}.
\subsubsection{Influence of $\mathrm{pH}$ on the nature of the
  interaction}
\begin{figure}[t!]
  \includegraphics[width=0.45\textwidth,
  trim={0.0cm 0.0cm 0cm 0.cm},clip]{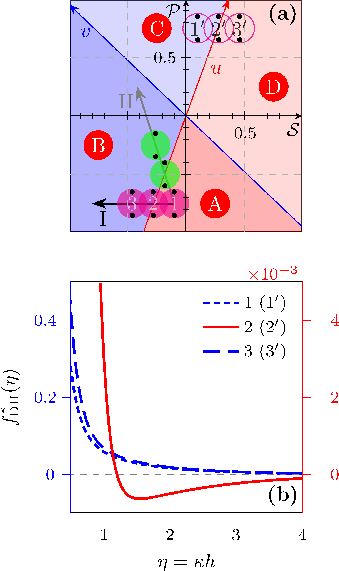}
  \caption{Influence of $\mathrm{pH}$ on the qualitative nature of the
    interaction between like-charged surfaces.  (a) Schematic
    representation of the influence of pH on the properties of
    ionizable anionic surfaces ($\sigc<0$) depicted on the $\Sc$-$\Pc$ plane
    ($\delta = 0.6$).  Circular regions represent the distribution of
    surface properties in an ensemble of suspended particles. Black
    points within the circle denote two representative particles in
    the ensemble, or surface regions of the particles, engaged in an
    interaction.  Increasing pH may be seen as a translation of the ensemble along a path of increasing fixed charge $\Sc$
    (path I or II), and drives a regime change in
    interactions. Process $1\to 2$ represents a transition from
    \textit{symmetric} to \textit{antisymmetric} interactions.
    Process $2 \to 3$ indicates a reversion back to the symmetric
    regime as the circular patch fully resides in region B. (b)
    Calculated interaction free energy profiles for flat-plate
    properties specified by the two black points, as the system
    properties translate along path I (process $1\to2\to3$). The
    profiles demonstrate the attraction at intermediate pH (red curve,
    stage 2) and re-entrant repulsion as a function of increasing pH
    (blue curves, stages 1 and 3), as observed in experiments. For
    $\kappa^{-1} = 10$ nm typical for the relevant experiment,
    $\Sc = \Pc= 0.5$ corresponds to a charge density
    $\sigc = 5.68\times10^{-3} e$ nm$^{-2}$. Similarly,
    $f_\tu{DH}^*(\eta)=1$ denotes a free energy per unit area of
    $5.68\times10^{-3} k_\tu{B}T$ nm$^{-2}$.  }
  \label{fig:pHEffect}
\end{figure}

The results obtained from the nonlocal theory, permit us to propose a
mechanism that may explain the experimentally observed effect of
$\mathrm{pH}$ on interactions between charged objects in aqueous and
non-aqueous media \cite{wang2024,wang2025}. In its most general form,
the influence of $\mathrm{pH}$ can be incorporated into our model by
introducing subsidiary relations which determine the surface charges
$\Sc$ and $\Pc$ as functions of the interfacial values of $\psi$ and
$\phi$, i.e., $\Sc = \Sc(\phi^\tu{s},\psi^\tu{s})$ and
$\Pc = \Pc(\phi^\tu{s},\psi^\tu{s})$. In PB theory, the charge
regulation boundary condition, which establishes a functional
correspondence between $\Sc$ and $\phi^\tu{s}$, is an example of this
type of relation \cite{ninham1971,Markovich2016}.  Using this concept
and the geometric point of view that we adopted earlier, variation in
$\mathrm{pH}$ may be seen as a process which translates the
coordinates $(\Sc, \Pc)$ of a charged surface along a particular curve
in $\Sc$-$\Pc$ plane. Technically, the shape of this curve is dictated
by the model which describes the ionization processes at the
liquid-solid interface. Since modeling of surface ionization is beyond
the scope of the current work, we illustrate our proposed mechanism
schematically by considering two slightly dissimilar surfaces whose
charge densities $p_1=(\Sc_1, \Pc_1)$ and $p_2=(\Sc_2, \Pc_2)$ are
located within a small circular patch in region A of the $\Sc$-$\Pc$
plane. The diameter of this zone qualitatively reflects the
distribution in particle or indeed surface properties, typical of
heterogeneous natural systems.  Figure \ref{fig:pHEffect}a depicts
this circular patch (circle 1) in the domain of $\sigc<0 $ and
$\sigp<0$, characteristic of an anionic system at low pH in an aqueous
electrolyte (Fig. \ref{fig:schm}c).  The two black points refer to
system properties $p_1$ and $p_2$, respectively. In this situation,
any two points (including $p_1$ and $p_2$) within the circular patch
produce a symmetric system which is characterized by purely repulsive
interactions. An increase in pH leads to an increase in fixed charge
density. We depict this by displacing the circle in the direction of
increasing $\lvert\sigc \vert$ along a curve that crosses the
$u$-axis. This permits antisymmetric interactions to arise, where,
based on the arguments presented in Sec. \ref{sec:DH}, a long range
attraction is inevitable. Figure \ref{fig:pHEffect}a displays two
examples (I and II) of such paths. In both cases, system interactions
exemplified by $p_1$ and $p_2$ (black points) transition from
repulsion to attraction (process $1 \to 2$). At a
given separation, a lower free
energy in an antisymmetric interaction compared to a symmetric case ensures that the former are thermodynamically favored
over the latter.  In real systems of particles that are capable of
sampling degrees of freedom such as
rotational and orientational modes, as well as permuting particle identity, antisymmetric interactions will be thermodynamically
preferred over symmetric interactions. With a further increase in pH,
the circular patch eventually reside entirely within the same region
or quadrant B (process $2 \to 3$). Now all interactions are purely
symmetric and therefore repulsive. Equivalent results are expected in
the domain of basic, positively charged particles interacting in
alcohols, where both $\sigc>0$ and $\sigp>0$
(Fig. \ref{fig:schm}c). Here, as pH decreases, the analogous process
unfolds in the diagonally opposite quadrant I of the $\Sc, \Pc$ plane
(process $1' \to 2' \to 3'$) (Fig. \ref{fig:pHEffect}a).

\subsection{The Sphere-Sphere Interaction}
Having established general considerations and trends underpinning
interactions between surfaces in a nonlocal medium, we now discuss
the calculation of interaction free energies between two
particles. The ability to calculate forces between particles enables
quantitative comparisons of the model with experimental
measurements. The force of interaction between two spheres separated
by a distance $\eta$ is determined by the free energy of
interaction $F(\eta)$ [Eq. \eqref{eq:F}]. We examine this free energy
for different systems, first by applying the Derjaguin approximation
(DA) to one-dimensional flat plate geometries, and subsequently using
numerical calculations for the two-sphere interaction in 2-dimensional
axisymmetric coordinates.

\subsubsection{Generalized Derjaguin Approximation}
In the context of the local PB theory, Derjaguin's approximation
relates the interaction force between two spherical particles to the
interaction free energy of the corresponding one-dimensional flat
plate geometry \cite{derjaguin1934}.
\begin{figure}[t!]
  \includegraphics[width=0.45\textwidth,
  trim={0.0cm -0.cm 0cm 0cm},clip]{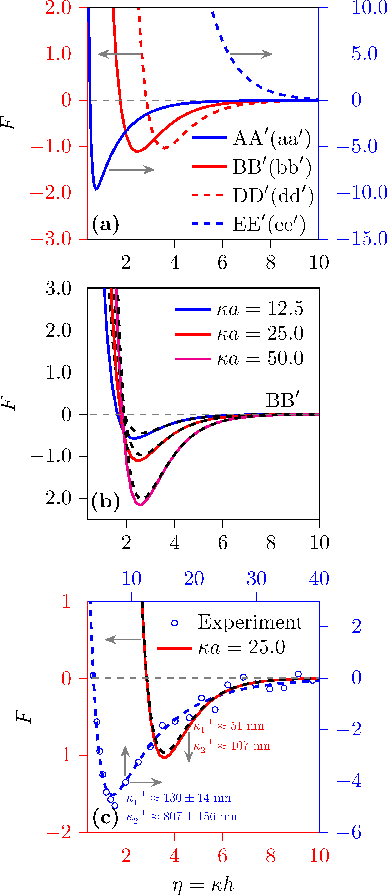}
  \caption{Free energy of interaction $F$ vs. intersurface separation $\eta$ for spheres. (a) $F(\eta)$ profiles for spheres of
    radius $a=\qty{2.5}{\micro\meter}$ ($\alpha = 25$) for
    systems specified by pairs of points in
    Fig. \ref{fig:phaseMap}.  (b) Effect of particle size on $F(\eta)$ for spheres characterized by the property pair
    $\mathrm{BB'}$: numerical calculations (solid lines) and the corresponding GDAs (dashed lines). In all calculations, $\delta = 0.4$,
    $\theta=16$, and salt concentration corresponds to $\kappa^{-1}=\qty{100}{\nano\meter}$. (c) Comparison of a calculated free energy profile (red curve, $\mathrm{DD}'$ in
    (a)) with an experimental measurement from Ref. \onlinecite{wang2026} (open symbols). In both cases, $\kappa_1^{-1}$ and $\kappa_2^{-1}$ are
    obtained by fitting the profiles to Eq. \eqref{eq:FEinfAsym2} (dashed curves).}
  \label{fig:FEcurves}
\end{figure}
The generalized Derjaguin Approximation (GDA) introduced by Schnitzer
and Morozov \cite{schnitzer2015} permits a calculation of the dimensional
interaction free
energy $F(\eta)$ of two spheres of dimensionless radius
$\alpha = \kappa a$ using the following integral
\begin{equation}
  \label{eq:GDA}
  F(\eta) = \frac{\alpha^2}{4 \tau}
  \int_\eta^\infty \frac{
    f^*(\zeta)}{2 \alpha +\zeta} \tu{d}\zeta, 
\end{equation}
where $f^*(\eta)$ is the dimensionless free energy of interaction
between two flat plates (see SI for detail). The above integral is
valid in the regime $\kappa a \gg 1$, which holds for large colloidal
particles in aqueous media. In general, $f^*(\eta)$ can be calculated
either by numerically solving the PB equation for a pair of flat
plates, or by using \eqref{eq:FE1dDHL} if the DH condition
$\phi \ll 1$ holds. When $\alpha \gg 1$ and $f^*(\eta)$ decays
rapidly, the integral in Eq. \eqref{eq:GDA} may be written as
\begin{equation}
\label{eq:DANond}
  F(\eta) = \frac{\alpha}{8 \tau}
  \int_\eta^\infty f^*(\zeta)\tu{d}\zeta,
\end{equation}
which, in the DH regime, has a closed form solution given by
\begin{equation}
\label{eq:FDHL_GDA}
  \begin{aligned}
    F_\tu{DHL}(\eta; \Sc_1, \Sc_2) = &
    \frac{\alpha}{8\tau}\Bigg[ 2 \Sc_1 \Sc_2 \ln \left(
    \frac{1+\exp(-\eta)}{1-\exp(-\eta)} \right) - \\
    & (\Sc_1^2+\Sc_2^2) \ln\left[1- \exp(-2\eta)\right]\Bigg].
  \end{aligned}
\end{equation}
In Sec. \ref{sec:DH}, we showed that in nonlocal theory the free
energy $f^*_\tu{DH}(\eta)$ can be view as the superposition of two
coupled local subsystems. Thus in the DH regime, Eq. \eqref{eq:DANond}
can be applied to each of the terms in Eq. \eqref{eq:Fe1dDH} provided
that the validity condition for the DA is satisfied for the longest
screening length of the problem given by $(\kappa \kappa_-)^{-1}$. For
the sphere-sphere interaction, this yields a formula analogous to
Eq. \eqref{eq:Fe1dDH}
\begin{equation}
 F_\tu{DH}(\eta) = \sum_{j\in \mathcal{A}} 
 \frac{F_\tu{DHL}(\kappa_j \eta; \Sc_{1j}^\tu{eff},
   \Sc_{2j}^\tu{eff})}{\kappa_j}
\end{equation}
which is valid when both DH and DA conditions are satisfied. When the
DH condition does not hold, we use Finite Element calculations in one
dimension to obtain $f^*(\eta)$ from Eq. \eqref{eq:fe1dEq}, and
subsequently evaluate the integral in Eq. \eqref{eq:GDA}
numerically. In the thin double layer regimes ($\delta \ll 1$ and
$\alpha \gg 1$) where axisymmetric numerical calculations of the free
energy are prohibitively costly, the above approach is fairly accurate
and economical.
\subsubsection{Numerical Calculations of the Interaction Free Energy}
Here we present results of numerical computations of free energies for two interacting spheres, and compare the obtained profiles both with results from the GDA as well as with experimental data. Figure \ref{fig:FEcurves} illustrates 
the dependence of $F$ on
intersurface separation $\eta$ for four different systems of
interacting spheres (Fig. \ref{fig:schm}a). In all calculations, we
assume that both particles have identical radii, $\alpha = 25$, but
their surfaces may carry different values of fixed and polarization
charge.  The salt concentration corresponds to a Debye length of $100$
nm, and the correlation parameter is $\delta =0.4$.  We examine
symmetric and antisymmetric systems by assigning the properties of the particles in each pair to different points on the $\Sc$-$\Pc$ plane in
Fig. \ref{fig:phaseMap}.

Pair-interaction systems characterized by points that are reflections of each other through the origin produce pair interaction profiles that are
indistinguishable. Accordingly, representative antisymmetric pairings
denoted by the interactions $\tu{A}\tu{A'} (\tu{a}\tu{a'})$,
$\tu{B}\tu{B'} (\tu{b}\tu{b'})$, and $\tu{D}\tu{D'} (\tu{d}\tu{d'})$
all exhibit nonmonotonic free energy profiles with a minimum in $F$ at
finite separation $\eta_\tu{min}$. Note that particles in each pairing
carry the same sign of fixed charge but the values of their
polarization charge differ.  Contrary to interactions in a local
medium, the interparticle force is attractive for $\eta>\eta_\tu{min}$
and turns repulsive for $\eta<\eta_\tu{min}$. On the other hand, for
symmetric systems such as $\tu{E}\tu{E'} (\tu{e}\tu{e'})$, the free
energy of interaction $F$ is a monotonically decreasing function of
$\eta$ implying a repulsive interaction at all separations.

Antisymmetric pairings of particles may produce a potential minimum of
significant depth at large separations.  A deep potential minimum
plays a decisive role in the appearance of clusters with large
interparticle separations in colloidal suspensions
\cite{wang2024,wang2025}.  Furthermore, the influence of particle size
on the magnitude of the attraction, and therefore implicitly, on the
depth of the minimum at long range is of particular experimental
interest \cite{wang2026}.  Figure \ref{fig:FEcurves}b displays the
numerically calculated interaction free energy for different particle
radii $\alpha$, and surface properties given by the pairing
$\tu{B}\tu{B'}(\tu{b}\tu{b'})$, together with the corresponding GDA
profiles computed using Eq. \eqref{eq:GDA}.  The agreement between the
GDA and the numerical calculations confirms the linear dependence of
$F$ on $\alpha$, i.e., $F \propto \alpha$ expected from
Eq. \eqref{eq:DANond}. However, at large intersurface separations
$\eta\gg\alpha$ we may expect a transition to a stronger dependence of
the interaction free energy on particle radius, i.e.,
$F \propto \alpha^2$, which may be relevant for experimental
observations \cite{wang2026}.

Figure \ref{fig:FEcurves}c compares a computed two-sphere interaction energy profile with a representative measurement for two silica particles reported in Ref. \onlinecite{wang2026}. Both profiles reveal minima at $\eta\approx5$ and inverse screening lengths $\kappa_1$ of the same order of magnitude of $\kappa$. However the screening length $\kappa_2^{-1}$ characterizing the decay of the experimentally measured attractive force is about an order of magnitude larger than the Debye length. Whilst the location of the minimum and the decay length of the repulsion are approximately captured, the decay length of the attraction is not captured within the present framework. It is likely that the model in its present form does not contain the physics required to explain the range of the experimentally measured electrosolvation attraction.

\begin{figure}[t!]
  \includegraphics[width=0.45\textwidth,
  trim={0.0cm -0.cm 0cm 0cm},clip]{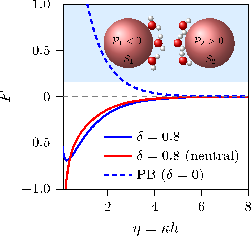}
  \caption{Free energy of interaction $F$ vs. intersurface separation $\eta$ for
    two spherical particles of size $\alpha = 2.5$ in the high salt
    regime ($a=2.5$ nm$, \kappa^{-1} = 1$ nm). Particles properties
    are characterized by a pair of points $p_1 = (-2.2, -1.7)$ and
    $p_2 = (-2.2, 1.7)$ in the $\Sc$-$\Pc$ plane, corresponding to
    negatively charged particles with a fixed charge density
    $\lvert \sigc \rvert=0.25e$ nm$^{-2}$ and carrying polarization
    charge density of opposite signs where $\lvert \sigp \rvert=0.2 e$
    nm$^{-2}$ (blue solid curve). Neutral spheres with $\sigc=0$ and
    $\sigp=\pm0.2 e$ nm$^{-2}$ display substantial attraction to
    contact (red curve). While PB theory in a local medium (dashed
    curve) predicts a purely repulsive interaction for the
    like-charged spheres, the nonlocal theory with a correlation
    parameter $\delta =0.8$ ($\xi=2$ \AA) and $\theta = 16$ envisages
    an attraction originating directly from the disparity in surface
    polarization.}
  \label{fig:FECurveHighSalt}
\end{figure}

Finally, we examine the interactions of nanometer scale objects in aqueous electrolytes containing high concentrations of monovalent salt ($100$ mM). Figure \ref{fig:FECurveHighSalt} displays calculations for like-charged and neutral spheres of radius $2.5$ nm, representative of a protein. Assuming a correlation 
length of $\xi=2$~\AA,  which is typical for bulk water, we find that substantial attraction is possible for spherical objects immersed in electrolytes containing physiological levels of salt. In contrast, PB theory in a local medium envisages repulsion for like-charged nanospheres in water, as expected.
 
\section{Conclusions}
A model of nonlocal electrostatics envisages richer
behavior and may offer greater insight into the mechanisms underpinning the
properties and interactions of particles and molecules in liquids. The
ability of interfacial solvent structuring to contribute to the
electrical potential on the same footing as fixed charge
immediately suggests an important role in electrokinetic phenomena
such as electrical mobility (Fig. \ref{fig:potDist}). Although we do
not compute mobilities in this work, the influence of interfacial
polarization on the sign and magnitude of the electrical potential at
the shear plane indicates significant impact on measured 
$\zeta$-potentials \cite{matyushov2024}. 

\begin{figure}[t!]
  \includegraphics[width=0.45\textwidth,
  trim={0.0cm -0.cm 0cm 0cm},clip]{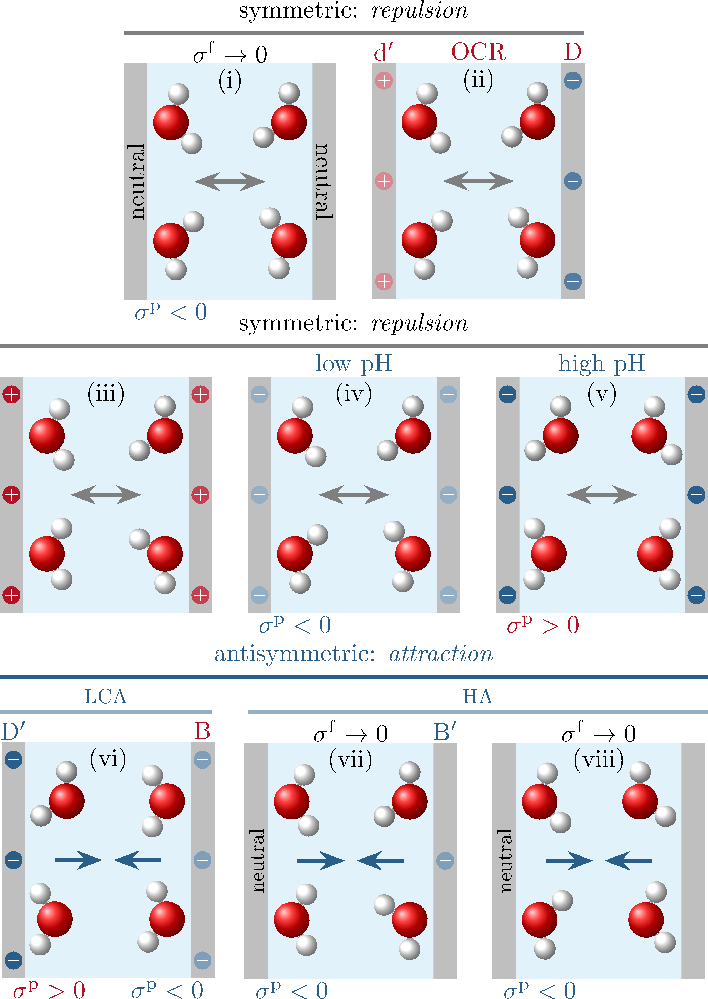}
  \caption{Possible scenarios and long-range interaction outcomes in
    water from the nonlocal model:
    like-charge attraction (LCA), hydrophobic attraction (HA) and
    opposite charge repulsion (OCR).  Symmetric interactions
    (repulsion): \textbf{(i)} Formally charge-neutral surfaces may
    repel solely on account of the electrical effects of interfacial
    polarization. \textbf{(ii, iv)} Weakly charged surfaces, of the
    same or opposite sign, may repel on account of a sufficiently
    strong interfacial polarization that dominates the surface charge
    (OCR: e.g., points $\tu{d}’$ and D in
    Fig. \ref{fig:phaseMap}). \textbf{(iii, v)} The electrical
    contribution from interfacial polarization reinforces that of the
    surface charge causing repulsion.  Note that scenarios (iv) and
    (v) could reflect interactions between anionic surfaces at low and
    high pH respectively. Antisymmetric interactions (attraction):
    \textbf{(vi)} Like-charged surfaces with dissimilar interfacial
    polarization (not necessarily opposite in sign) may appear
    oppositely charged in the far-field producing an attraction at
    large separations. \textbf{(vii, viii)} Possible scenarios
    underpinning long-ranged hydrophobic attraction (HA): either a
    small imbalance in charge (left), or polarizations of opposite
    sign (right), on two predominantly neutral surfaces could be
    sufficient to generate an attraction.  }
  \label{fig:expPatterns}
\end{figure}

Figure
\ref{fig:expPatterns} displays a qualitative overview of
representative outcomes for the long-range component of interactions
in water, under selected conditions. Major highlights include the
ability of anionic, charge-neutral or near-neutral surfaces to either
attract or repel. Implicitly, cationic surfaces in aqueous media
always repel. Significantly, the model anticipates attraction between net neutral objects, driven by interfacial polarization, which has been previously alluded to as a mechanism behind the long-range hydrophobic attraction \cite{despa2007}. The nonlocal framework also points to the possibility of counterintuitive long-range
repulsion between oppositely charged surfaces. 

Overall, the present model contains the ingredients required to qualitatively
capture the experimentally observed sign-dependent attraction or
repulsion between like-charged spheres in water and aqueous solvents,
thus furnishing a possible mechanistic picture underpinning the
experimentally observed electrosolvation force 
\cite{wang2024,wang2025,wang2026}. In particular, depending on the
parameter values chosen, the model readily reflects the appearance of
non-monotonic potentials that are attractive at long range, and
display minima of depth $\approx 1 k_{\tu{B}}T$ located at
dimensionless separations as large as $\kappa h \approx 5$.  The
interactions turn repulsive at shorter range in line with the
experiments. Figure \ref{fig:pHEffect} also illustrates qualitatively
the ability of the model to capture the experimentally observed role
of pH in controlling the formation of clusters in particles capable of
charge regulation \cite{wang2024,wang2025,wang2026}.

However, the main points of quantitative departure between the present
model and experimental observations to date concern the depth of the
attractive minima, as well as the extremely long range of the
attraction seen in experiments. Experiments display a long-ranged
minimum of substantial depth ($\gtrsim 5 k_{\tu{B}}T$) in the
interaction of like-charged particles (see
Fig. \ref{fig:FEcurves}c). Next, for correlation lengths
$\xi\lesssim 10$ nm, the present model anticipates a decay length for
the attraction that is slightly larger than the Debye length, i.e.,
$\kappa_2^{-1}\gtrsim\kappa^{-1}$. The decay length for the repulsion
in turn, is expected to be a factor of two smaller:
$\kappa^{-1}_1=0.5\kappa_2^{-1}$. Experiments however consistently
reveal an attraction characterized by a decay length
$\kappa_2^{-1}\gg\kappa^{-1}$ and a repulsive interaction at shorter
separation characterized by a decay length of
$\kappa^{-1}_{1}\gtrsim\kappa^{-1}$. In addition, the calculations
presented here for microspheres interacting under low salt conditions
entail values of fixed and polarization charge densities that are
about one to two orders of magnitude smaller than those expected from
experimental estimates and MD simulations respectively.

Furthermore, interaction energy profiles calculated for the
microspheres in Fig. \ref{fig:FEcurves} assume a correlation length
$\xi=10$ nm in water. This value is admittedly much larger than both
the molecular scale of $\approx 3$ \AA~and a typical correlation
length of $1$-$3$ \AA~used in nonlocal models of water
\cite{kornyshev1984,rubashkin2014}.  However non-linear spectroscopy
measurements have reported correlation lengths in water ranging from a
few nanometers to as high as $20-25$ nm
\cite{shelton2012,shelton2014,chen2016,duboisset2018,dedic2019}.
Indeed the magnitude of the relevant correlation length remains a
subject of ongoing discussion \cite{jungwirth2018}.

Importantly, there have been suggestions that the correlation length
relevant to interactions is not necessarily the value characterizing
pure bulk water, but may in fact strongly depend on contributions from
surface structural wavelengths which are properties of the interacting
objects themselves. This points to the possibility of a variety of length scales underpinning interactions in fluids that depend both on the properties of the
objects and that of the medium \cite{kornyshev1989, leikin1990,
  schlaich2024}.  A natural conclusion is then that interaction decay
lengths measured in experiments are likely to be highly
non-universal. It is worth noting parenthetically that
non-universality in measured decay length scales has in fact been
reported in recent experiments \cite{wang2026}. All the above
considerations comparing the results of the nonlocal model and
experiments on colloidal microspheres indicate important additional
physics at play in the experimental interaction that is not captured
within the present model, setting the stage for future conceptual
advances.

Calculations for neutral and like-charged molecular-scale matter
immersed in electrolytes containing physiologically relevant
concentrations of salt have furnished further interesting
insight. Assuming a correlation length of $\xi=2$ \AA~ typical for
bulk water, as well as reasonable values for the fixed and
polarization charge densities, we find that nanometer scale objects
can experience non-negligible (attractive) pair interaction energies
of $\approx -0.5 k_{\tu{B}}T$ purely on account of interfacial
polarization (Fig. \ref{fig:FECurveHighSalt}). This result echoes an
early report on the impact of built-in surface dipoles in the nonlocal
interaction of lipid bilayers \cite{belaya1987,belaya1994b}. Our results
show that a difference in interfacial polarization is sufficient to
generate an attraction between both neutral as well as electrically
like-charged objects at the molecular scale in solution. Although in
the calculation presented here, the attractive pair interaction energy
is not substantially larger than the thermal energy scale, the
strength of this interaction would be sufficient to form a separate
phase owing to additive multivalent interactions that occur in
clusters of molecules as discussed in Ref. \onlinecite{wang2024}. This
result suggests that solvent structuring at the molecular interface
could provide a rather general attractive force capable of driving
condensation, cluster formation, and intramolecular collapse in
biological systems, e.g., chromatin condensation, biological phase
segregation and possibly even in protein folding \cite{franzmann2018,
  polymenidou2018, wadsworth2023}.

This study takes a step in the direction of formulating a
self-consistent theory of interactions of objects in fluids. We have
focused in particular on the nature of the interaction at larger
separations because energy barriers and potential minima at larger
interparticle distances can prove decisive in determining kinetically
controlled outcomes in reactions, organization, assembly and structure
formation in the fluid phase.  Furthermore, although there is great
interest in interactions in water, experiments demonstrate that
attraction between and cluster formation amongst like-charged objects
in fluids are general phenomena, observed across a range of solvents
in which the molecular length scale and correlation lengths can be
substantially larger than in water \cite{shelton2012}.  Overall it
appears that there are general unifying principles associated with
solvent-governed interactions in fluids, and that apparently highly
anomalous features observed in experiment may be qualitatively
captured within a comparatively simple model of interactions in a
nonlocal medium.

\section*{Supplementary Material}
Detailed derivations of the governing equations have
been provided in the Supplementary Material. The Supplementary
Material also contains a Mathematica application which allows
interactive exploration of interaction regimes for two parallel
plates.

\begin{acknowledgements}
  The authors gratefully acknowledge funding from the European
  Research Council (ERC) under Horizon Europe (No. 101199352).\\
\end{acknowledgements}
\section*{Author Declarations}
\noindent
\textbf{Conflict of Interest}

The authors have no conflicts to disclose.

\section*{data availability}
All data are available within the article or supplemental information.

\appendix
\section{Solution of Nonlocal Governing Equations for a Flat Plate
  Geometry in DH regime}
\label{sec:appPot}
The solution of the one-dimensional form of Eqs. \eqref{eq:nondPsi}
and \eqref{eq:nondPhi} over a finite segment $[0, \eta]$ can be
represented in closed form using analytical techniques
\cite{paillusson2010,behjatian2025}. For simplicity, we first consider
a completely neutral flat plate at $x=\eta$ interacting with a flat
plate carrying surface charges $\Sc$ and $\Pc$ at $x=0$. Accordingly,
the potential fields due to the flat plate located at $x=0$ are given
by
\begin{equation}
\label{eq:phi1dSngl}
  \Phi(x;\Sc, \Pc) = \sum_{i \in \mathcal{A}}
  \frac{\kappa_i \gamma_i(\Sc, \Pc)}{\sinh\kappa_i\eta}
  \cosh[\kappa_i(x-\eta)]    
\end{equation}
and
\begin{equation}
\label{eq:psi1dSngl}
  \Psi(x;\Sc, \Pc) = \sum_{i\in \mathcal{A}}
  \frac{\gamma_i(\Sc, \Pc)}{\kappa_i\sinh\kappa_i\eta}
  \cosh[\kappa_i(x-\eta)]
\end{equation}
where 
\begin{equation}
\label{eq:gamma}
\gamma_i(\Sc, \Pc) = \frac{(\Sc+\Pc)\theta -\Sc
  \kappa_j^2}{\kappa_i^2-\kappa_j^2}
\end{equation}
is a function of two variables, and $\mathcal{A}=\{+,-\}$.  We
emphasize that in Eq. \eqref{eq:gamma} the indices
$i,j \in \mathcal{A}$ and $i \ne j$.  Utilizing the linearity and the
principle of superposition in DH regime, the solutions for the general
case of two charged walls separated by a distance $\eta$
(Fig. \ref{fig:schm}) can be represented by
\begin{equation}
\label{eq:phiDH2p}
\phi^\tu{DH} (x) = \Phi(x;\Sc_1, \Pc_1)+
\Phi(\eta-x; \Sc_2, \Pc_2),
\end{equation}
and
\begin{equation}
\label{eq:psiDH2p}
\psi^\tu{DH} (x) = \Psi(x;\Sc_1, \Pc_1)+
\Psi(\eta-x; \Sc_2, \Pc_2),
\end{equation}
respectively. We note that in the limit $\eta \to \infty$,
Eqs. \eqref{eq:phi1dSngl} and \eqref{eq:psi1dSngl} recover the
solution for a semi-infinite domain
\begin{equation}
     \Phi(x;\Sc, \Pc) = \sum_{i \in \mathcal{A}}
    \kappa_i \gamma_i(\Sc, \Pc) \exp(-\kappa_i x),
\end{equation}
\begin{equation}
    \Psi(x;\Sc, \Pc) = \sum_{i\in \mathcal{A}}
    \frac{\gamma_i(\Sc, \Pc)}{\kappa_i}
    \exp(-\kappa_i x).
\end{equation}
Accordingly, the coefficients $A_\pm$ given in Sec.  \ref{sec:1dpot}
can be expressed as
\begin{equation}
    A_i = \kappa_i \gamma_i(\Sc, \Pc).
\end{equation}

\section{Free Energy of Interaction in the DH-Regime}
\label{sec:appEffCharge}
Using Eqs. \eqref{eq:phiDH2p} and \eqref{eq:psiDH2p}, the equilibrium
free energy per unit area of two flat plates in DH regime can be
calculated from Eq. \eqref{eq:fe1dEqDH}. Accordingly, the free energy
of interaction per unit area can be expressed as
\begin{equation}
  \begin{aligned}
  f_\tu{DH}(\eta) = & \frac{1}{8 \pi A_0}
  \Bigg[
     B_+ (\coth \kappa_+ \eta -1)+
     \frac{C_+}{\sinh \kappa_+ \eta}
     +  \\
     & B_-( \coth \kappa_-\eta -1 )+
     \frac{C_-}{\sinh \kappa_-\eta} 
       \Bigg] 
  \end{aligned}
\end{equation}
where $B_+ = \Omega_+^2 (u_1^2+u_2^2)$, $C_+ = 2 \Omega_+^2 u_1 u_2$,
$B_- = \Omega_-^2 (v_1^2+v_2^2)$, $C_- = 2 \Omega_-^2 v_1 v_2$, are
constants which are described by the natural coordinates of each plate
$u_i = \Pc_i + n \Sc_i$ and $v_i = \Pc_i - m \Sc_i$ in $u$-$v$
plane. Here, $\Omega_\pm$ are two positive constants which are given
by
\begin{equation}
  \Omega_+ = \frac{\theta(\kappa_+^2-1)}
  {\sqrt{\kappa_+(\theta-1)
      (\kappa_+^4-2\kappa_+^2+\theta)}},
\end{equation}
and
\begin{equation}
  \Omega_- = \frac{\theta}
  {\sqrt{\kappa_-(\kappa_+^4-2\kappa_+^2+\theta)}}.
\end{equation}
Defining the effective charges
$(\Sc_{1+}^\tu{eff}, \Sc_{2+}^\tu{eff})$ and
$(\Sc_{1-}^\tu{eff}, \Sc_{2-}^\tu{eff})$, the nonlocal interaction of
two flat plates in the DH regime can be recast as a superposition of
interactions between two local subsystems, each characterized by the
dimensionless screening lengths $\kappa_+$ and $\kappa_-$,
respectively. In other words, defining
$f^*_\tu{DH}(\eta) = 8 \pi A_0 f_\tu{DH}(\eta)$, we may rewrite the
interaction free energy as
\begin{equation}
    f^*_\tu{DH}(\eta) = \sum_{j \in \mathcal{A}}   
    f^*_\tu{DHL}(\kappa_j \eta; \Sc_{1j}^\tu{eff},
    \Sc_{2j}^\tu{eff})
\end{equation}
where $\mathcal{A}=\{+,-\}$, and the effective surface charges are
expressed as $\Sc_{i+}^\tu{eff} = \Omega_+ u_i$ and
$\Sc_{i-}^\tu{eff} = \Omega_- v_i$, respectively.\\

\bibliography{ref.bib}

\end{document}